\documentclass[12pt,preprint]{aastex}
\newcommand{\SB}{\mbox{${\rm MJy}\,{\rm sr}^{-1}$}}

\newcommand{\eg}{e.g.,~}
\newcommand{\ie}{i.e.,~}
\newcommand{\etal}{et al.~}

\newcommand{\ISO}{{\it ISO}}
\newcommand{\IRAS}{{\it IRAS}}
\newcommand{\Spitzer}{{\it Spitzer}}
\begin{document}

\title{Absolute Calibration and Characterization of the Multiband Imaging Photometer for Spitzer: 
       IV. The Spectral Energy Distribution Mode}
\author{Nanyao Lu\altaffilmark{1}, 
	Paul S. Smith\altaffilmark{2}, 
        Charles W. Engelbracht\altaffilmark{2}, 
        Alberto Noriega-Crespo\altaffilmark{1}, 
        Jane Morrison\altaffilmark{2}, 
        Karl D. Gordon\altaffilmark{2}, 
        John Stansberry\altaffilmark{2}, 
        Francine R. Marleau\altaffilmark{1}, 
        George H. Rieke\altaffilmark{2}, 
        Roberta Paladini\altaffilmark{1}, 
        Deborah L. Padgett\altaffilmark{1}, 
	Jocelyn Keene\altaffilmark{1}, 
        William B. Latter\altaffilmark{1}, 
        Dario Fadda\altaffilmark{1},
	Jeonghee Rho\altaffilmark{1} 
        }

\altaffiltext{1}{\Spitzer\ Space Telescope Science Center, 
                 MS 314-6,
                 California Institute of Technology, Pasadena, CA 91125;
                 lu@ipac.caltech.edu}
\altaffiltext{2}{Steward Observatory, The University of Arizona, Tucson, AZ 85721}

\date{\center{(Accepted to PASP, January 22, 2008)}}

\begin{abstract}
\indent

The Spectral Energy Distribution (SED) mode of the Multiband Imaging 
Photometer for {\it Spitzer\/} (MIPS) {\it Space Telescope\/}
provides low-spectral resolution 
(R $\approx$ 15-25) spectroscopy in the far infrared using the MIPS 
70~\micron\ detector.  A reflective grating provides a dispersion of 
1.7~\micron\ per pixel, and an effective wavelength coverage of
52.8--98.7~\micron\ over detector rows 1-27.  The final 5 detector rows
are contaminated by second-order diffracted light and are left 
uncalibrated.   The flux calibration is based on observations of 
MIPS calibration stars with 70~\micron\ flux densities of 0.5--15$\,$Jy. 
The point-source flux calibration 
accuracy is estimated to be 10\% or better down to about 0.5$\,$Jy 
at the blue end of the spectrum and to $\sim 2\,$Jy near the red end.
With additional uncertainties from the illumination and aperture 
corrections included, the surface brightness calibration of extended 
sources is accurate to $\sim 15\%$.  Repeatability of better 
than 5\% is found for the SED mode through multiple measurements
of several calibration stars.

\end{abstract}


\section{Introduction} \label{sec1}

The Spectral Energy Distribution (SED) mode of the Multiband 
Imaging Photometer for \Spitzer\ (MIPS; Rieke \etal 2004) 
aboard the {\it Spitzer Space Telescope} (\Spitzer; Werner 
\etal 2004) provides a 
capability for obtaining long-slit, low-resolution (R $\approx$
15--25) spectra in the far infrared (53--98~\micron).  This extends 
\Spitzer's spectroscopic wavelength coverage to beyond that of the higher-resolution 
Infrared Spectrograph (Houck \etal 2004) that operates up to 40~\micron.
The SED slit is two detector pixels ($\sim 20\arcsec$) in width,
and 12 pixels ($\sim 2\arcmin$) in length where the full wavelength
coverage is available.  An inoperative detector module restricts 
the wavelength coverage to only 67-98~\micron\ over the last 
4 columns of the detector array of 16 columns times 32 rows.

Observations with \IRAS\ and the {\it Infrared Space Observatory} (\ISO) 
have shown that most starburst galaxies and active galactic nuclei 
have their far-infrared dust emission peaking between 50 and 100~\micron.  
The MIPS SED mode is well tailored to capture a spectral section 
around the peak of the dust emission, and thus can provide a good 
constraint on the temperatures and masses of the bulk of dust particles 
in these energetic systems.  Although it has a lower spectral 
resolution than the \ISO\ Long-Wavelength Spectrometer (LWS), the MIPS 
SED mode is more sensitive (\eg at 60~\micron, the 5-$\sigma$ continuum
sensitivity of 82$\,$mJy in a 500-sec on-target integration achieved
in the MIPS SED mode for point sources is roughly 3 times that
of the \ISO\ LWS integrated over the wavelength coverage 
of each MIPS SED pixel; see Gry \etal 2003).  It also provides
more spatial information and 
has demonstrated an improved degree of repeatability.  
The MIPS SED mode has been used by \Spitzer\ observers
for various projects,  ranging from determining galaxy spectral 
energy distributions in the far-infrared (\eg Kennicutt \etal 2003)
to characterizing circumstellar dust emission (\eg Low \etal 2005).

Using the same detector and internal stimulators (STIM), the SED mode 
shares many calibration characteristics (\eg dark current) with the 
70~\micron\ imaging photometry mode.  The calibration of the latter is 
described in 
detail by Gordon \etal (2007; hereafter GEF07).  In this paper, we 
first give a brief overview of the operation and calibration principles
of the MIPS SED mode (\S2) and then describe in more detail specific
calibration
issues, including illumination correction (\S3), 
wavelength calibration (\S4), aperture corrections (\S5), and flux 
calibration (\S6).  A summary is given in \S7.

\section{Operation and Calibration Principles of the MIPS SED Mode} \label{sec2}

The astronomical observing template of the SED mode, as illustrated in 
Fig.~1, provides 
pairs of data frames between the target position (ON) and a nearby 
sky position (OFF).  The MIPS scan mirror is used to chop between ON 
and OFF, and an observer can choose chop throws of +3\arcmin, 
+2\arcmin, +1\arcmin, or $-$1\arcmin\ on the sky.  
The observer can request either 
a pointed observation or an MxN raster map of a larger region
(with M, N between 1 and 100). 
For a pointed observation (or at each raster position in the case 
of a map), there is a basic set of 6 pairs of ON and OFF frames.
The first 3 pairs of data-collection events (DCEs), each of $\sim$10~s 
(or 3~s if observing very bright sources) in duration,
are obtained with the target 
placed near the center of detector column 10 (hereafter referred to as
dither position 1) and the next 3 pairs with the target near the center
of column 5 (dither position 2).
The telescope is moved to position the target on the dither positions.
Bracketing each of the 3 pairs of DCEs are STIM exposures to track detector
responsivity variations.
Each STIM is preceded by a DCE at OFF.
The observer repeats this basic 
set by specifying a number of cycles ($N_c$) in order to reach 
the desired signal-to-noise ratio (S/N). 
For a mapping observation, the value of $N_c$ 
remains the same for all the raster positions.

The basic instrumental calibration of the SED mode is similar to 
that of the 70~\micron\ imaging mode and follows eqs.~(4) to (16) in 
Gordon \etal (2005; hereafter GRE05).  The single difference is that 
in the SED  mode, the detector Y-axis coincides with the spectral 
dispersion axis.  As a result, an illumination correction (IC) 
frame obtained from some sky position in the SED mode contains a 
wavelength-dependent factor, $Z$, inherited from the intrinsic 
spectral signature of the sky emission (cf. \S3).  Eq.~(15) of GRE05 
can be modified for the SED mode as follows:
\begin{eqnarray}
   U_{illum}(i,j) = Z(j)\,O(i,j)/S(i,j), 
\end{eqnarray}
where $i$ and $j$ are respectively the detector column and row indices, 
$O(i,j)$ is the combined optical response of the telescope and 
instrument, and $S(i,j)$ is the STIM illumination pattern.  Since 
$j$ correlates linearly with wavelength (cf. \S4), we have expressed
$Z$ as a function of $j$ in eq.~(1).

The responsivity-corrected, dark-subtracted and IC-corrected data
frame of eq.~(16) in GRE05 becomes
\begin{eqnarray}
   U_{data}(i,j) = I(i,j)/Z(j),
\end{eqnarray}
where $I(i,j)$ is the spectral image of the sky projected onto 
the detector array.

The subsequent flux calibration of eq.~(2) converts MIPS (instrumental) 
units to Jy.  This is accomplished via observations of 
standard stars (cf. \S6), 
from which we derive a mean nominal spectral response function, $R$, 
for the SED mode:
\begin{equation}
R^{-1}(j) = f_{\nu}(j)C^{-1}_{aper}(j) / (\sum^N_{i=1}\,I_{sub}(i,j) / Z(j)),
\end{equation}
where the summation represents the spectral extraction within a pre-defined 
aperture of N columns times 1 row, $f_{\nu}(j)$ is the known stellar flux 
density at the central wavelength of detector row $j$, $C^{-1}_{aper}(j)$
is the corresponding aperture correction factor so that $f_{\nu}(j)C^{-1}_{aper}(j)$
is the net flux density within the chosen aperture, and $I_{sub}(i,j)$ is 
the sky-subtracted signal from the target star.  The nominal spectral
response differs slightly from the conventionally defined spectral response 
function, $R_0$, commonly seen in the literature: $R(j)\,Z(j) = 
R_0(j)$.  Since $Z(j)$ appears in both eqs.~(2) and (3), it is 
removed from the data during flux calibration.

Calibration observations of the SED mode
were reduced using the MIPS Data Analysis Tool
(DAT, v2.96, see GRE05). The results presented here are derived
from either the co-added images at a given dither position or a mosaic 
image from both dither positions.  The image pixel scale is always kept
equal to the detector pixel size of 9.8\arcsec\  $\times$ 9.8\arcsec.

\section{Illumination Correction} \label{sec3}

\subsection{Observational Results} \label{sec3.1}

The characterization of the combined effects of a non-uniform illumination
pattern and the difference in illumination between sky and the undispersed 
STIM flashes can be calibrated by imaging celestial 
sources of uniform surface brightness (GRE05).
The Zodiacal light is generally too faint for deriving such an IC 
for the SED mode. Instead, we have utilized SED 
observations of diffuse Galactic emission 
(with \IRAS\ $I_{\nu}(60\,\micron)/I_{\nu}(100\,\micron) 
\sim 0.2-0.3$) near the Galactic plane.  Once an adequate number 
of independent observations have been acquired, median filtering 
is performed to exclude any spatial structure that might be present 
in individual fields.  To reach an adequate signal level (\ie $\sim$5--10 
times the level of the dark current), we targeted regions where \IRAS\ 
$I_{\nu}(60\,\micron) = 200-300\,\SB$ and where no \IRAS\ point sources 
are identified within a radius of $\sim$5\arcmin--10\arcmin. Each of 
the selected regions is observed in a 4x1 map with a chop throw of 
2\arcmin.

Individual observations
were examined for signs of ``bad'' sky positions (\eg a point source 
accidentally in the slit), and a subset was chosen for median filtering
to generate a final IC image.  As an illustration, Fig. 2 compares 
the individual IC observations by plotting their normalized mean spatial
profiles, together with the resulting median.

Fig.~2 reveals that the instrumental sensitivity drops substantially
for the first couple of detector columns.  A possible explanation could 
be that the end of the slit vignettes these detector columns.  The first
detector column also shows a much greater pixel-to-pixel noise level 
compared with the other detector columns.  This entire column is masked 
out in the S17 and later versions of the MIPS SED mode data reduction 
pipeline.

\subsection{Residual Systematics} \label{sec3.2}

The adopted approach to constructing the IC
is quite efficient, reaching high signal levels
over the entire detector array in a reasonable amount of telescope 
time. However, there are some residual systematic variations of up to 
$\sim 15\%$ along the spatial axis of the slit. 
This effect is illustrated
in Fig.~3 using the data from a 16x1 mapping observation of the point 
source-like galaxy Arp$\,$231.   The target was scanned along the slit 
with a step size that matches the detector pixel size.  For each raster
pointing, the ON frames from one of the two default dither positions
have been co-added and sky subtracted, and a 3-column aperture was used 
to extract a spectrum.  The results from the map are shown
in Fig.~3 for two detector row ranges.

Fig.~3 reveals two important characteristics: (1) Excluding the noisy
first detector column (which affects the first two data points in 
each plot), 
there is a flux variation of up to 10-15\% from column to column. (2) This 
column-to-column variation has only a weak dependence on wavelength.
Since the amplitude of this residual systematic variation
appears to scale with incident 
flux, a first-order empirical correction can be made by averaging
a number of observations similar to that of Arp$\,$231.  Table~1 lists
a simple column-wise correction derived from two such mapping observations.
Dividing the values in
Table~1 into the IC image in a column-wise way leads
to a composite IC that reduces the residual flux variation to $\sim 5\%$ 
(except for the first detector column) and brings the flux 
measured between the two default dither positions into reasonable agreement 
for the observed calibration stars.  This composite IC is used throughout
this paper.

\section{Wavelength Calibration} \label{sec4}

The formal dispersion solution was based on two observations of the bright 
planetary nebula NGC$\,$6543 that shows prominent [\ion{N}{3}]57.330~\micron\ and 
[\ion{O}{3}]88.356~\micron\ emission lines.  The two observations are identified by 
Astronomical Observation Request (AOR) tags of 13437952 and 20719616, respectively.

Each observation of NGC~6543 was reduced independently.
The sky-subtracted, mosaic images from the two observations
were averaged together; a spectrum was subsequently extracted using a 
5-column aperture and is shown in Fig.~4.    The MIPS
instrument design yields, to first order, a linear dispersion and
the instrumental optical model suggests
that the bandpass of each detector pixel only varies
by $\pm 1.5\%$ from the nominal value.  Assuming uniform dispersion,
the measured
line positions yield a dispersion of 1.70~\micron\ per
pixel and a wavelength of 
53.66~\micron\ for the center of the first detector row at the blue end of 
the spectrum.

Each line fitting uses 4 independent data points; the wavelength calibration 
uncertainty is thus estimated to be on the order of 0.4~\micron\ ($\approx 
0.5*1.7{\rm ~\micron}/\sqrt{4}$) for uniform dispersion.  
Differences in line positions among various line profile fitting methods 
(\eg centroiding using moments) are all within this uncertainty 
estimate. 
For a given centroiding method,
the differences in the emission-line positions between the two observations
of NGC~6543, obtained about 19 months apart, are on the order of 0.1~\micron.

The effective wavelength coverage over detector rows 1-27 of the SED mode
is 52.8--98.7~\micron\ and the wavelength assignments for each detector row
are listed in column (2) of Table~2.  The remaining 5 detector rows at 
wavelengths $> 98$~\micron\ are contaminated by shorter wavelength, second-order
light and are left uncalibrated.

\section{Aperture Corrections} \label{sec5}

In Fig. 5, the observed mean spatial profiles at 70 and 90~\micron\ are shown
in comparison with the predicted point spread functions (PSF) from 
the \Spitzer\ Tiny Tim model (Krist 2002).
The measured profiles were derived from the average
of 9 observations of 3 bright stars through MIPS Campaign (MC) 24 
in August 2005.
Averaging was independently done at each dither 
position to test if the PSF has a dependence on array 
position.   We averaged detector rows 10 and 11 to produce
the spatial profile at 70~\micron.  Detector rows 22 and 23 were averaged 
for an estimate of the PSF at 90~\micron. The abscissa of Fig.~5 is the
distance along the slit length, measured from the corresponding profile 
peak.  Because of a small spectral tilt, the profile center location on 
the detector array has a small dependence on wavelength. 
We smoothed the fine-sampled Tiny Tim model image with a square 
box of width D~= 1, 1.3, and 1.7 times the detector pixel size, and
derived a predicted PSF by integrating the smoothed image over the slit
width of 2 pixels.  All of the model profiles were normalized so that 
they yield the same intensity sum from the 3 sampled positions closest
to the profile peak at dither position 1; the measured profile 
from dither position 2 was then normalized in a similar way with 
respect to the model profile of D~= 1.

The measured PSF for the SED mode matches the Tiny Tim models very well 
at small radii.  For radii that encompass the first Airy ring, the measurements
agree best with the D~= 1 model curve, although the sparse sampling 
does not allow for the D~= 1.3 model to be robustly ruled out.
At large positive distances (45\arcsec--60\arcsec), the observed profile 
appears to be broader than the model predictions.  However, the excess 
has a significance of only 2--2.5$\,\sigma$.  Better agreement 
with the model predictions is seen at large negative distances.
The magnitude of the observed excess, its asymmetry,
and the fact that it is seen at both dither positions make it unlikely 
that this effect is caused by either IC residual systematics or detector 
flux nonlinearity.

In columns (3)--(6) of Table~2, we have calculated the aperture correction 
factors (C) using the D~= 1.0 model for extraction aperture sizes of 2--5
detector columns {\it centered} on an unresolved target.   
The difference in aperture corrections between D $=1.0$ and D $=1.3$ is 
less than 2\% at any wavelength.   The aperture correction ratio between 
the 3- and 5-column aperture sizes in Table~2 can also be measured 
using the bright calibration stars and the result is compared in Fig.~6 
with various model predictions.   It is evident that the data agree 
best with the D~= 1 case, and are within 2--3\% of the model predictions.

About 10\% of the flux lies outside the radius of 50\arcsec\ in the case 
of D~= 1.  In contrast, the measured PSF is roughly twice as bright as 
the D~= 1 model prediction at radii $\gtrsim 50$\arcsec.  This implies
that the flux calibration for an extended source using the aperture 
correction factors in Table~2 could be off by up to 10\% if this "light 
excess" at large radii originates from the PSF core and affects a large
fraction of the PSF disk.

\section{Flux Calibration} \label{sec6}

\subsection{Calibration Stars} \label{sec6.1}

The primary flux calibrators for the SED mode are moderate to bright 
stars from a list compiled for the flux calibration of the MIPS 
70~\micron\ photometric mode (GEF07).
Their effective photospheric temperatures, $T_e$, and model flux 
densities at 71.42~\micron, $f_{\nu}(71.42)$, are given 
in Engelbracht \etal (2007), where 71.42~\micron\ is the effective
wavelength of MIPS broad-band 70~\micron\ 
photometric system.  The model spectrum of a star over the spectral
range of the SED mode is simply represented as a Planck function with 
$T = T_e$ anchored at an adopted $f_{\nu}(71.42)$, which 
can be either the model prediction from Engelbracht \etal (2007) or 
the MIPS 70~\micron\ photometric observation from GEF07.
Note that for stars, the color correction necessary to convert a 
MIPS 70~\micron\ photometric measurement to a monochromatic flux
density is unity.

Because the effective sensitivity of the SED mode on stars decreases 
steeply as wavelength increases, we have chosen calibration stars
in three flux ranges of $f_{\nu}(71.42)$: (1) 3 stars 
brighter than 10$\,$Jy, (2) 8 moderately bright stars with 
$f_{\nu}(71.42) = 2$--6$\,$Jy,  and (3) 11 ``faint" stars 
with 0.45~Jy~$< f_{\nu}(71.42) <$ 2~Jy.   Observations were 
made (except during a couple of early MIPS campaigns) so that S/N $ 
\gtrsim 10$ was reached for stars in (1) over the entire spectrum, 
for stars in (2) shortward of $\sim 80$~\micron, and for those 
in (3) shortward of 65-70~\micron.   This tiered strategy 
allows for reasonable integration times at all three flux levels, 
and it tests whether there is a significant flux nonlinearity in 
the detector responsivity.

Table~3 lists observations of the calibration stars through August 
2005 (\ie MC24).  The table columns are: (1) the star catalog name,
(2) the spectral type,  (3) the model $f_{\nu}(71.42)$ from 
Engelbracht \etal (2007) with the flux uncertainty given in 
the parentheses,  (4) the MIPS-measured $f_{\nu}(71.42)$ by GEF07 
using their PSF fitting method, (5) a $f_{\nu}(71.42)$ derived 
from the SED-mode measurement using the aperture correction
factors in Table~(2) and the spectral response function in 
Table~(4), of which the derivation is explained in \S6.2, (6)
the S/N ratio of the SED-mode measurement,  (7) the number of cycles
executed in the SED observation, (8) the MIPS campaign number, and 
(9) the AOR tag in the \Spitzer\ data archive.  The uncertainty for 
the flux densities in column~(4) is taken to be 5\% (see GEF07).

A number of the calibration stars in Table~3 have been observed in multiple 
campaigns.  These include stars in all three flux level categories.  
The SED-mode results for these stars (when measurement S/N ratios are high 
enough, \eg S/N $>$ 15) indicate a stable instrument performance with time.
Repeatability of $\sim 2\%$ and $\lesssim 5$\% were measured at the blue
and red ends of the spectrum, respectively.

\subsection{Nominal Spectral Response Function} \label{sec6.2}

For each observation of a calibration star, the spectrum was 
extracted from the sky-subtracted mosaic image using a 5-column
aperture.  If a star was observed multiple times, the resulting
spectra were averaged.

Fig.~7 compares 71.42~\micron\ flux density ratios between 
(i) the SED-mode, (ii) the photospheric model predictions 
and (iii) MIPS 70~\micron\ PSF-fitted photometry for the calibration
stars.  For the purpose of these figures, the SED-mode flux scale 
was simply derived by using a flux conversion factor that results in 
the median flux density ratio of (ii) to 
(i) $\approx 1$ for stars with $f_{\nu}(71.42) > 1$~Jy.
While there is no apparent flux dependency in the ratio between 
the model predictions and the SED-mode results, the three brightest 
stars are significantly below 1 in both Fig.~7(a) and (b).  These 
stars are at the high end of the flux range calibrated for the MIPS 
photometric system (GEF07), and therefore, their 70~\micron\ 
photometric fluxes may suffer some moderate flux nonlinearity.  
Since these stars are crucial in the SED-mode flux calibration 
because they are the only measurements having high S/N at 
the red end of the spectrum,  we chose to base the SED-mode flux
calibration on the photospheric model predictions.  The model
predictions are consistent with the MIPS 70~\micron\ photometric
fluxes below $\sim 10\,$Jy.

The SED-mode spectrum of each calibration star was aperture 
corrected and divided into the adopted stellar model spectrum
to yield 
the inverse of the nominal spectral response function, $R^{-1}$, 
in units of Jy per MIPS unit, where $R\/$ is as defined in 
eq.~(3).  We used an inverse noise-squared weighting to 
determine the average $R^{-1}$.  In addition, 
at a given wavelength, a star is rejected if the $S/N < 10$.

Table~4 gives the resulting mean nominal spectral response 
function in the standard \Spitzer\ units of \SB\ per MIPS 
unit, where 1~Jy~= 443~\SB\ $\times (9.8\arcsec\ \times 
9.8\arcsec)$.   The results in Table~4 also reflect the MIPS
data reduction pipeline convention that has the IC normalized 
by its median within the image section [2:16,1:31].  
The quoted uncertainty in $R^{-1}$ is the sample standard 
deviation of the mean, $\sigma_{\rm mean}$, given by 
\begin{equation}
\sigma^2_{\rm mean} = 
{1 \over (\Sigma\,w_i)^2/\Sigma\,w^2_i - 1}({\Sigma\,w_ir^2_i \over \Sigma\,w_i} - <r>^2),
\end{equation}
where $r_i$ and $w_i$ are respectively the response and weight 
from the $i$th sample star, and $<r>$ is the weighted mean response.
Finally, the actual number of stars averaged at each wavelength 
is given in column~(5).

As discussed in \S2, the nominal spectral response function 
includes the spectral signature of the diffuse Galactic emission 
via the IC used in the data reduction.  Thus, $R^{-1}$ {\it 
changes} if a {\it different} IC is used in the data reduction.

\subsection{Absolute Flux Accuracy and Cross Comparisons} \label{sec6.3}

Flux or wavelength-dependent systematic errors in the SED-mode
flux calibration can be checked by comparison with model 
predictions for the calibration stars at various representative
wavelengths.  Such a comparison at 71.42~\micron\ is displayed 
in Fig.~7(c).  In Fig.~8, similar plots are shown for $\lambda 
\approx 60$, $75$ and $90$~\micron.   In these plots, the Planck
function model spectrum for each star is normalized to 
the MIPS 70~\micron\ photometric flux density.   As expected, 
the SED-mode flux densities have fairly large uncertainties
at both 75~\micron\ and 90~\micron\ for stars with $f_{\nu}(71.42)
< 2$~Jy.  In spite of this, the flux density ratios do not 
appear to have any obvious dependency on flux at the three 
selected wavelengths outside of the three brightest stars 
that may have their MIPS 70~\micron\ photometric flux densities 
underestimated by 5-10\%.   A check of the uncertainty in 
the determination of the flux calibration is offered by 
the scatter in the data points plotted in Fig.~8.   The scatter 
is $< 10$\% at 60~\micron\  down to at least 0.5~Jy and down 
to $\sim$2~Jy at 90~\micron.

Comparisons with the results from past spacecraft suggest 
that the absolute flux calibration of the SED mode is accurate
to 10\% or better for point sources.   For example, Fig.~9 
compares the MIPS SED-mode spectra of the galaxies NGC~4418 
and Mrk$\,$231 with $IRAS\/$ measurements at 60 and 100~\micron.  
Also shown in Fig.~9 is an $ISO\/$ LWS spectrum of Mrk~231.   
Generally, the data agree to within 10\%. These comparisons 
also suggest that flux nonlinearity for the SED mode is 
insignificant over a 70\micron\ flux density range of 
0.5--40$\,$Jy.

\subsection{Extended-Source Flux Calibration} \label{sec6.4}

Since the SED-mode flux calibration is based on stars, uncertainties
in the aperture correction and the IC systematic effect described in \S3.2
have little impact on the calibration of a point source.
However, these do introduce
additional uncertainty in the surface brightness calibration of an 
extended source. 
Although difficult to quantify because of the lack of available SED-mode
observations of well-calibrated extended sources,
Table~5 summarizes an estimated error budget for this case.
The total uncertainty in flux density is estimated to 
be on the order of 15\% or less.

The galaxy merger system Arp$\,$299
offers an opportunity to evaluate the performance of the MIPS SED mode
for an extended source.
It
was observed using a 4x1 raster map (AOR 
$=$ 12919296)
centered at R.A. $=11^h28^m32.3^s$
and Dec $= 58\arcdeg33\arcmin43\arcsec$ (J2000.0) and with the slit oriented at 
a position angle of $119\arcdeg$. 
A raster step 
size equal to the width of the SED slit was used so that the map covers the entire 
optical extent of the system with no spatial gaps.  Fig.~10 compares 
an \ISO\ LWS spectrum of Arp$\,$299 with the spatially integrated MIPS
spectrum within a rectangular aperture of $78.4\arcsec \times 88.2\arcsec$. 
This aperture size is slightly larger than the circular LWS beam whose 
diameter ranges from 
$84.6\arcsec$ to $77.2\arcsec$ over 50--100~\micron\ (Gry \etal 2003).
The overall agreement between the {\sl ISO\/} and {\it Spitzer\/} spectra
is better than 15\%.

\section{Summary} \label{sec7}

We have described in this paper the calibration of the SED mode
of MIPS, based on the current calibration status of the instrument
(as of \Spitzer\ Pipeline version S17).  Our main points are: 
(1) The SED optical system is stable, has a dispersion of 
1.7~\micron\ per detector pixel, and covers 52.8--98.7~\micron\ 
over detector rows 1--27.  
(2) The final 5 detector rows are contaminated by the second-order 
diffracted light, and are not calibrated.  (3) The first detector 
column should be excluded in any subsequent analysis of SED-mode 
data.  (4) The illumination correction has a column-wise residual
variation on the order of 5\%.  (5) The observed PSF shows some 
excessive light and asymmetry at radii $> 50\arcsec$ when compared 
with the prediction from \Spitzer\ Tiny Tim model.  This adds 
an additional uncertainty of up to 10\% in the flux calibration 
of extended sources.  (6) Point-source flux calibration is accurate
to 10\% or better, down to $\sim 0.5\,$Jy at the blue end of 
the spectrum, and to $\sim$2~Jy near the red end.  (7) No significant 
flux nonlinearity is seen over a 70~\micron\ flux density range 
of 0.5--40$\,$Jy.  (8) The corresponding surface brightness accuracy
for extended sources is estimated to be $\lesssim 15\%$ due to 
additional uncertainties in the illumination and aperture corrections.

\acknowledgments

This work is based on observations made with the {\it Spitzer Space
Telescope}, which is  operated by the Jet Propulsion Laboratory, 
California Institute of Technology under NASA contract 1407. This
work was also supported by the MIPS IT contract 1255094 to 
the University of Arizona.  P. S. Smith acknowledges support from 
JPL contract 1256424 to the University of Arizona.



%

\newpage


%
\begin{deluxetable}{rcrc}
\tabletypesize{\scriptsize}
\tablenum{1}
\tablewidth{0pt}
\tablecaption{Column-wise Division Factors to IC}
\tablehead{
\colhead{Det.~Col.}   & \colhead{Correction} & \colhead{Det.~Col.}   & \colhead{Correction}}
\startdata
1 \phn\phn  & 0.529  &   9\phn\phn  & 1.008   \\
2 \phn\phn  & 1.102  &  10\phn\phn  & 1.082   \\
3 \phn\phn  & 0.995  &  11\phn\phn  & 1.026   \\
4 \phn\phn  & 1.099  &  12\phn\phn  & 1.052   \\
5 \phn\phn  & 0.938  &  13\phn\phn  & 0.958   \\
6 \phn\phn  & 0.919  &  14\phn\phn  & 1.045   \\
7 \phn\phn  & 0.956  &  15\phn\phn  & 1.098   \\
8 \phn\phn  & 0.948  &  16\phn\phn  & 1.005   \\
\enddata
\end{deluxetable}
\newpage

%
\begin{deluxetable}{rccccc}
\tabletypesize{\scriptsize}
\tablenum{2}
\tablewidth{0pt}
\tablecaption{Aperture Corrections$^a$}
\tablehead{
\colhead{Det.~Row}   & \colhead{$\lambda(\micron)$} & \colhead{C(2cols)} & \colhead{C(3cols)} 
                         & \colhead{C(4cols)} & \colhead{C(5cols)} \\
\colhead{(1)}   & \colhead{(2)} & \colhead{(3)} & \colhead{(4)} & \colhead{(5)} & \colhead{(6)}}
\startdata
 1\phn\phn\phn\phn &  53.66   & 1.931   &1.763  &1.673      & 1.591\\
 2\phn\phn\phn\phn &  55.36   & 1.961   &1.785  &1.701      & 1.616\\
 3\phn\phn\phn\phn &  57.06   & 1.991   &1.807  &1.729      & 1.642\\
 4\phn\phn\phn\phn &  58.76   & 2.022   &1.829  &1.757      & 1.668\\
 5\phn\phn\phn\phn &  60.46   & 2.055   &1.851  &1.785      & 1.695\\
 6\phn\phn\phn\phn &  62.16   & 2.095   &1.874  &1.809      & 1.724\\
 7\phn\phn\phn\phn &  63.86   & 2.134   &1.896  &1.833      & 1.752\\
 8\phn\phn\phn\phn &  65.56   & 2.174   &1.918  &1.857      & 1.781\\
 9\phn\phn\phn\phn &  67.26   & 2.214   &1.941  &1.881      & 1.810\\
10\phn\phn\phn\phn &  68.96   & 2.254   &1.963  &1.905      & 1.838\\
11\phn\phn\phn\phn &  70.66   & 2.297   &1.988  &1.929      & 1.867\\
12\phn\phn\phn\phn &  72.36   & 2.347   &2.016  &1.951      & 1.894\\
13\phn\phn\phn\phn &  74.06   & 2.397   &2.044  &1.974      & 1.921\\
14\phn\phn\phn\phn &  75.76   & 2.447   &2.072  &1.997      & 1.948\\
15\phn\phn\phn\phn &  77.46   & 2.497   &2.100  &2.019      & 1.975\\
16\phn\phn\phn\phn &  79.16   & 2.547   &2.128  &2.042      & 2.003\\
17\phn\phn\phn\phn &  80.86   & 2.602   &2.160  &2.065      & 2.029\\
18\phn\phn\phn\phn &  82.56   & 2.662   &2.194  &2.090      & 2.054\\
19\phn\phn\phn\phn &  84.26   & 2.721   &2.229  &2.114      & 2.079\\
20\phn\phn\phn\phn &  85.96   & 2.781   &2.264  &2.139      & 2.104\\
21\phn\phn\phn\phn &  87.66   & 2.840   &2.298  &2.163      & 2.129\\
22\phn\phn\phn\phn &  89.36   & 2.900   &2.333  &2.188      & 2.154\\
23\phn\phn\phn\phn &  91.06   & 2.965   &2.372  &2.215      & 2.179\\
24\phn\phn\phn\phn &  92.76   & 3.033   &2.414  &2.243      & 2.204\\
25\phn\phn\phn\phn &  94.46   & 3.102   &2.455  &2.272      & 2.229\\
26\phn\phn\phn\phn &  96.16   & 3.170   &2.497  &2.300      & 2.254\\
27\phn\phn\phn\phn &  97.86   & 3.239   &2.539  &2.329      & 2.279\\
\enddata
\tablenotetext{a}{Derived from the \Spitzer\ Tiny Tim model smoothed by
	          a square box of a width equal to the detector pixel size.}
\end{deluxetable}
\newpage

\begin{deluxetable}{clrrrrrrr}
\tabletypesize{\footnotesize}
\tablenum{3}
\tablewidth{0pt}
\tablecaption{Observations of SED Calibration Stars}
\tablehead{
\colhead{Star}   & \colhead{Type} & \colhead{$f_{mod}$(unc.)$^a$} & \colhead{\ $f_{\rm MIPS}^b$}
                 & \colhead{$f_{\rm SED}^c$}  & \colhead{(S/N)$_{\rm SED}^d$}   & \colhead{N$_{c}$}  & \colhead{MC} &  \colhead{AOR} \\
                 &                    & \colhead{(Jy)}  & \colhead{(Jy)}  & \colhead{(Jy)}\\
\colhead{(1)}    & \colhead{(2)}      & \colhead{(3)}   & \colhead{(4)}  & \colhead{(5)}
                 & \colhead{(6)}      & \colhead{(7)}   & \colhead{(8)}  & \colhead{(9)}}
\startdata
HD006860  &    M0III    &  5.33(0.20) &  5.36  &   5.47\phn\phn   &  49\phn\phn\phn\phn   &   9   &  24  &   15811840 \\
HD018884  &    M1.5III  &  4.64(0.16) &  4.61  &   4.96\phn\phn   &  28\phn\phn\phn\phn   &   6   &  18  &   13124352 \\
HD029139  &    K5III    & 12.84(0.45) & 11.41  &  13.13\phn\phn   &  73\phn\phn\phn\phn   &   6   &  19  &   13309440 \\
HD045348  &    F0II     &  3.08(0.07) &  2.94  &   2.96\phn\phn   &  22\phn\phn\phn\phn   &   9   &  18  &   13124608 \\
          &             &             &        &   2.95\phn\phn   &  22\phn\phn\phn\phn   &   9   &  19  &   13309952 \\
          &             &             &        &   2.98\phn\phn   &  22\phn\phn\phn\phn   &   9   &  21  &   13615872 \\
HD051799  &    M1III    &  0.59(0.02) &  0.61  &   0.70\phn\phn   &   7\phn\phn\phn\phn   &  12   &  19  &   13310208 \\
HD060522  &    M0III    &  0.75(0.02) &  0.73  &   0.80\phn\phn   &   8\phn\phn\phn\phn   &  12   &  20  &   13440256 \\
HD062509  &    K0IIIb   &  2.61(0.08) &  2.62  &   2.65\phn\phn   &  25\phn\phn\phn\phn   &   9   &  20  &   13440000 \\
HD070272  &    K4.5III  &  0.55(0.02) & ...... &   0.59\phn\phn   &   6\phn\phn\phn\phn   &  12   &  20  &   13440512 \\
HD081797  &    K3II-III &  2.89(0.10) &  2.97  &   2.88\phn\phn   &  23\phn\phn\phn\phn   &   9   &  21  &   13616384 \\
HD082308  &    K5III    &  0.54(0.02) &  0.51  &   0.49\phn\phn   &   5\phn\phn\phn\phn   &  12   &  21  &   13616640 \\
HD082668  &    K5III 	&  1.42(0.20) &  1.49  &   1.30\phn\phn   &   6\phn\phn\phn\phn   &   6   &   7  &    9655552 \\
HD091056  &    M0III    &  0.45(0.10) & ...... &   0.78\phn\phn   &   6\phn\phn\phn\phn   &  12   &  18  &   13124864 \\
          &             &             &        &   0.52\phn\phn   &   4\phn\phn\phn\phn   &  12   &  19  &   13310464 \\
          &             &             &        &   0.63\phn\phn   &   6\phn\phn\phn\phn   &  18   &  23  &   15423488 \\
HD108903  &    M3.5III  & 17.00(1.77) & 16.16  &  19.08\phn\phn   &  91\phn\phn\phn\phn   &   6   &  18  &   13123840 \\
          &             &             &        &  18.77\phn\phn   &  89\phn\phn\phn\phn   &   6   &  19  &   13309184 \\
          &             &             &        &  19.00\phn\phn   &  90\phn\phn\phn\phn   &   6   &  22  &   15248640 \\
          &             &             &        &  18.98\phn\phn   &  90\phn\phn\phn\phn   &   6   &  23  &   15422720 \\
HD120477  &    K5.5III  &  0.67(0.02) &......  &   0.78\phn\phn   &   7\phn\phn\phn\phn   &  12   &  18  &   13125120 \\
          &             &             &        &   0.68\phn\phn   &   7\phn\phn\phn\phn   &  15   &  22  &   15249408 \\
HD123123  &    K2III    &  0.49(0.02) &  0.44  &   0.48\phn\phn   &   6\phn\phn\phn\phn   &  18   &  23  &   15423744 \\
HD124897  &    K1.5III  & 14.34(0.78) & 13.60  &  14.32\phn\phn   &  69\phn\phn\phn\phn   &   6   &  10  &   11625216 \\
          &             &             &        &  14.67\phn\phn   &  70\phn\phn\phn\phn   &   6   &  18  &   13124096 \\
          &             &             &        &  14.55\phn\phn   &  70\phn\phn\phn\phn   &   6   &  22  &   15248896 \\
          &             &             &        &  14.63\phn\phn   &  70\phn\phn\phn\phn   &   6   &  23  &   15422976 \\
HD131873  &    K4III    &  3.36(0.12) &  3.31  &   3.32\phn\phn   &  28\phn\phn\phn\phn   &   9   &  21  &   13616128 \\
HD137759  &    K2III    &  0.50(0.02) & ...... &   0.63\phn\phn   &   6\phn\phn\phn\phn   &  15   &  22  &   15249664 \\
HD150798  &    K2II-III &  2.99(0.09) & ...... &   2.92\phn\phn   &  23\phn\phn\phn\phn   &   9   &  20  &   13439744 \\
HD164058  &    K5III    &  3.31(0.10) &  3.38  &   3.32\phn\phn   &  25\phn\phn\phn\phn   &   9   &  19  &   13309696 \\
          &             &             &        &   3.22\phn\phn   &  24\phn\phn\phn\phn   &   9   &  20  &   13439232 \\
          &             &             &        &   3.25\phn\phn   &  24\phn\phn\phn\phn   &   9   &  21  &   13615616 \\
          &             &             &        &   3.19\phn\phn   &  24\phn\phn\phn\phn   &   9   &  22  &   15249152 \\
          &             &             &        &   3.17\phn\phn   &  24\phn\phn\phn\phn   &   9   &  23  &   15423232 \\
HD198542  &    M0III    &  0.82(0.04) &  0.81  &   0.86\phn\phn   &   8\phn\phn\phn\phn   &  12   &  21  &   13616896 \\
HD211416  &    K3III 	&  1.27(0.05) & ...... &   1.30\phn\phn   &   6\phn\phn\phn\phn   &   6   &   7  &    9657856 \\
\enddata
\tablenotetext{a}{Model-predicted flux density at 71.42\micron\ from Engelbracht \etal 
(2007) with the uncertainty given in the parentheses.}
\tablenotetext{b}{MIPS photometric flux density at 71.42\micron\ from GEF07.}
\tablenotetext{c}{SED-mode measured flux density at 71.42\micron\ (see the text).}
\tablenotetext{d}{S/N ratio of the SED-mode measurement.}
\end{deluxetable}
\newpage

%
\begin{deluxetable}{rcrrr}
\tabletypesize{\scriptsize}
\tablenum{4}
\tablewidth{0pt}
\tablecaption{Mean Inverse Nominal Spectral Response Function}
\tablehead{
\colhead{Row}   & \colhead{$\lambda$} & \colhead{($R^{-1}$)$^a$} & \colhead{$\sigma_{\rm mean}^b$} & \colhead{Stars}\\
                & \colhead{(\micron)} & \colhead{(\SB/MIPS)} & & \\
\colhead{(1)}   & \colhead{(2)} & \colhead{(3)} & \colhead{(4)} & \colhead{(5)}}
\startdata
 1\phn\phn  &  53.66  &   7695\phn\phn\phn\phn\phn\phn &  108\phn\phn   & 19\phn \\
 2\phn\phn  &  55.36  &   8440\phn\phn\phn\phn\phn\phn &  126\phn\phn   & 19\phn \\
 3\phn\phn  &  57.06  &   9695\phn\phn\phn\phn\phn\phn &  128\phn\phn   & 19\phn \\
 4\phn\phn  &  58.76  &  10576\phn\phn\phn\phn\phn\phn &  153\phn\phn   & 19\phn \\
 5\phn\phn  &  60.46  &  11453\phn\phn\phn\phn\phn\phn &  142\phn\phn   & 19\phn \\
 6\phn\phn  &  62.16  &  12775\phn\phn\phn\phn\phn\phn &  159\phn\phn   & 19\phn \\
 7\phn\phn  &  63.86  &  14194\phn\phn\phn\phn\phn\phn &  195\phn\phn   & 17\phn \\
 8\phn\phn  &  65.56  &  14792\phn\phn\phn\phn\phn\phn &  216\phn\phn   & 15\phn \\
 9\phn\phn  &  67.26  &  16524\phn\phn\phn\phn\phn\phn &  314\phn\phn   & 13\phn \\
10\phn\phn  &  68.96  &  17293\phn\phn\phn\phn\phn\phn &  221\phn\phn   & 12\phn \\
11\phn\phn  &  70.66  &  19031\phn\phn\phn\phn\phn\phn &  227\phn\phn   & 10\phn \\
12\phn\phn  &  72.36  &  20141\phn\phn\phn\phn\phn\phn &  230\phn\phn   & 10\phn \\
13\phn\phn  &  74.06  &  21327\phn\phn\phn\phn\phn\phn &  288\phn\phn   & 10\phn \\
14\phn\phn  &  75.76  &  22620\phn\phn\phn\phn\phn\phn &  329\phn\phn   & 10\phn \\
15\phn\phn  &  77.46  &  23606\phn\phn\phn\phn\phn\phn &  352\phn\phn   & 10\phn \\
16\phn\phn  &  79.16  &  24616\phn\phn\phn\phn\phn\phn &  228\phn\phn   & 10\phn \\
17\phn\phn  &  80.86  &  27105\phn\phn\phn\phn\phn\phn &  423\phn\phn   & 10\phn \\
18\phn\phn  &  82.56  &  27845\phn\phn\phn\phn\phn\phn &  402\phn\phn   & 10\phn \\
19\phn\phn  &  84.26  &  29469\phn\phn\phn\phn\phn\phn &  618\phn\phn   &  7\phn \\
20\phn\phn  &  85.96  &  31197\phn\phn\phn\phn\phn\phn &  429\phn\phn   &  6\phn \\
21\phn\phn  &  87.66  &  33085\phn\phn\phn\phn\phn\phn &  683\phn\phn   &  5\phn \\
22\phn\phn  &  89.36  &  34751\phn\phn\phn\phn\phn\phn &  771\phn\phn   &  5\phn \\
23\phn\phn  &  91.06  &  35385\phn\phn\phn\phn\phn\phn &  750\phn\phn   &  5\phn \\
24\phn\phn  &  92.76  &  35992\phn\phn\phn\phn\phn\phn &  497\phn\phn   &  5\phn \\
25\phn\phn  &  94.46  &  38665\phn\phn\phn\phn\phn\phn &  821\phn\phn   &  4\phn \\
26\phn\phn  &  96.16  &  38647\phn\phn\phn\phn\phn\phn & 1671\phn\phn   &  3\phn \\
27\phn\phn  &  97.86  &  39349\phn\phn\phn\phn\phn\phn & 1010\phn\phn   &  3\phn \\
\enddata				
\tablenotetext{a}{$R^{-1}$ is as defined in eq.~(3) in the text with the aperture correction
factor from Table~2.}
\tablenotetext{b}{Sample standard deviation of the mean as given in eq.~(4) in the text.}
\end{deluxetable}
\newpage

%
\begin{deluxetable}{lr}
\tablenum{5}
\tablewidth{0pt}
\tablecaption{Error Budget for Flux Calibration of Extended Sources}
\tablehead{
\colhead{Error Source}   & \colhead{Estimated Error}}
\startdata
        Point-source flux uncertainty        &  $\le  10\%$ \\
        Residual IC uncertainty:   	     &  $\sim  5\%$ \\
	Aperture correction uncertainty      &  $<    10\%$ \\
      -------------------------------------------  & ------------ \\
	Total$^a$:			     &  $<    15\%$ \\
\enddata
\tablenotetext{a}{The total error squared equals the quadratic sum 
of the three individual errors.}
\end{deluxetable}
\newpage
\pagebreak


\begin{figure}
\plotone{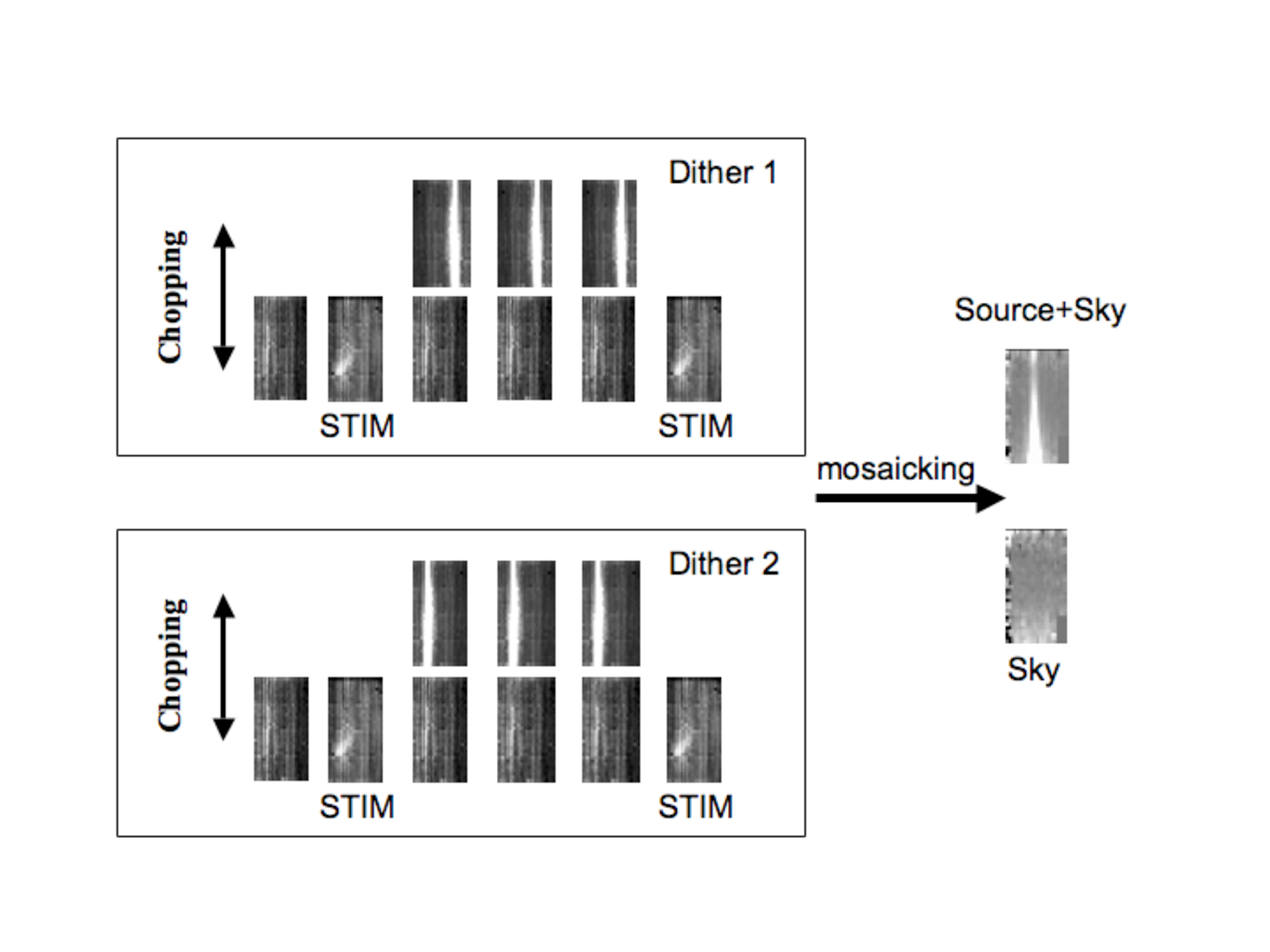}
\caption{
     A sketch of the basic set of frames taken with the astronomical observing template 
     of the MIPS/SED mode. The sequence of the observation is from left to right.  
     Dithers are carried out by simple spacecraft offset, while the chopping is carried
     out by the Cryogenic Scan Mirror Mechanism.  For both dither positions, STIM frames
     are always taken at the background position. Final images of the target plus background
     and background alone are produced by combining appropriate individual frames in 
     mosaic.}
\end{figure}
\vspace{0.3in}

\begin{figure}
\plotone{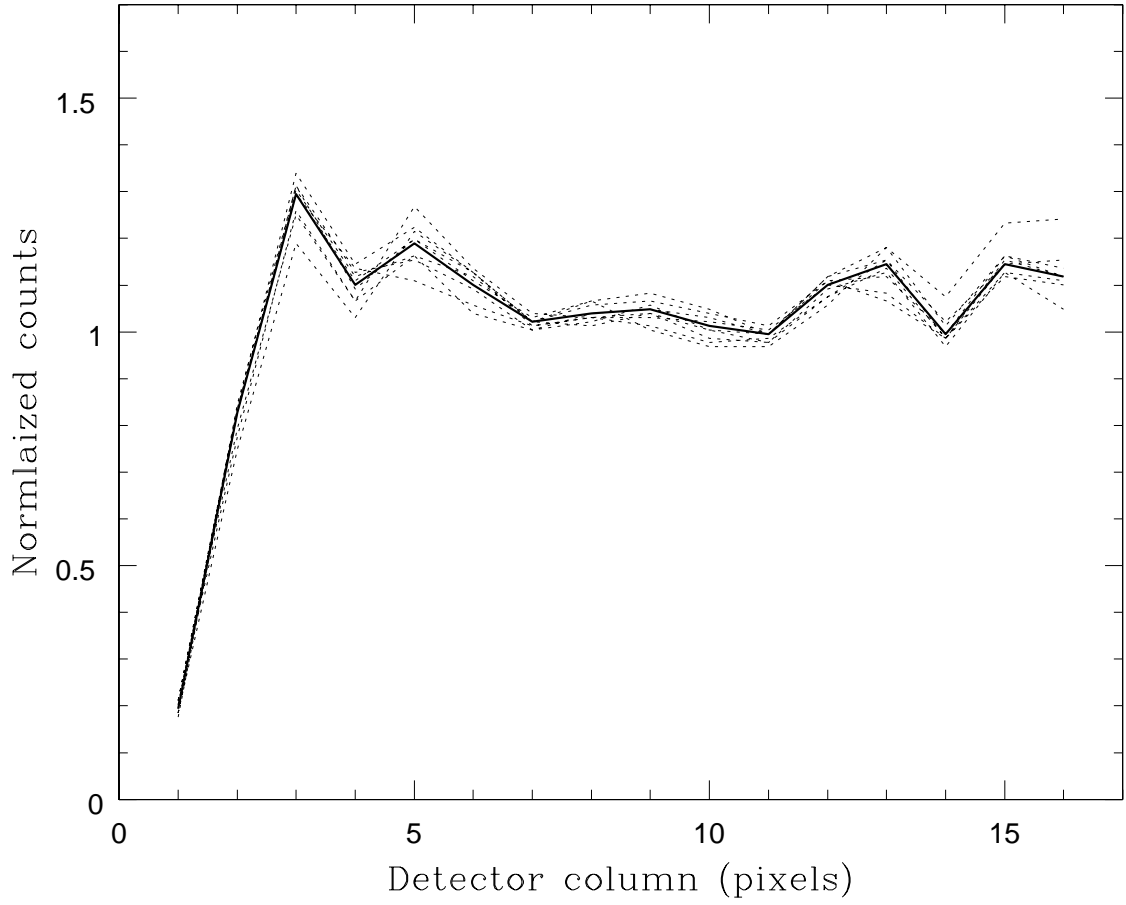}
\vspace{-0.7in}
\caption{
     Plot of the average spatial profile (dotted lines) over detector rows 9-15 
     for each of the 9 IC observations acquired between MIPS campaigns 20 
     and 24.  The thick solid curve shows the median result.}
\end{figure}
\vspace{0.3in}

\begin{figure}
\plotone{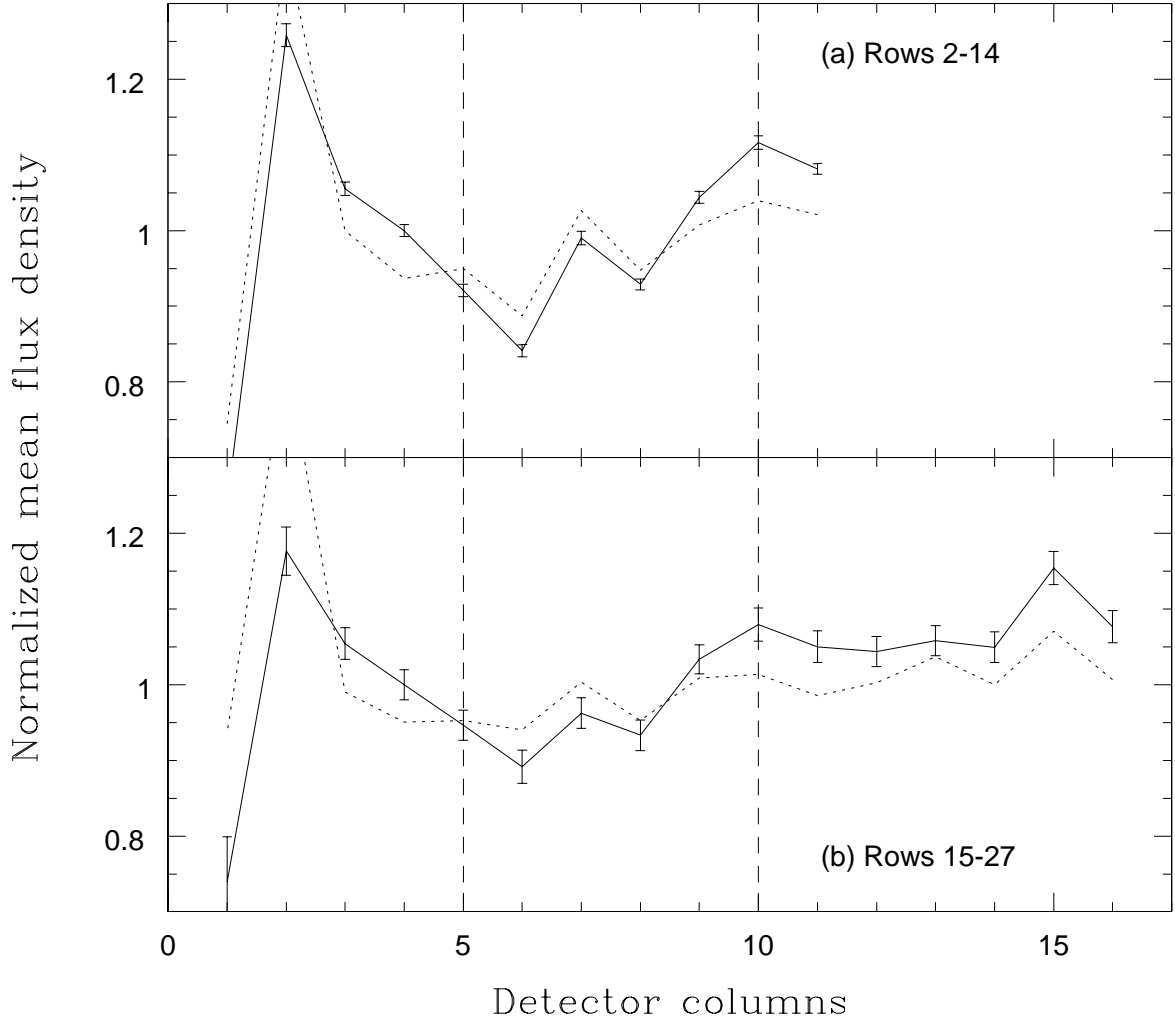}
\caption{
     Plots of the observed average signal (solid lines) from Arp~231
     using a 3-column aperture at each of the 16 \Spitzer\ positions
     of a raster-map observation.  Panel (a) is averaged over detector
     rows 2-14 (55--75~\micron) and (b) over rows 15-27 (77-97~\micron).
     A column-wise correction (see \S3.2 and Table~1) yields the dotted
     curves.  The two dashed vertical lines indicate 
     the default dither positions.  The truncation of the data in 
     panel (a) beyond column 11 is due to the inoperative detector 
     module.Note that the first two flux points in each 
     plot are affected by the first detector column, which is much 
     noisier than the other columns. Excluding these first two data
     points, the dotted curve in each plot shows a variation of $\sim 
     5$\%.
     }
\end{figure}

\begin{figure}
\plotone{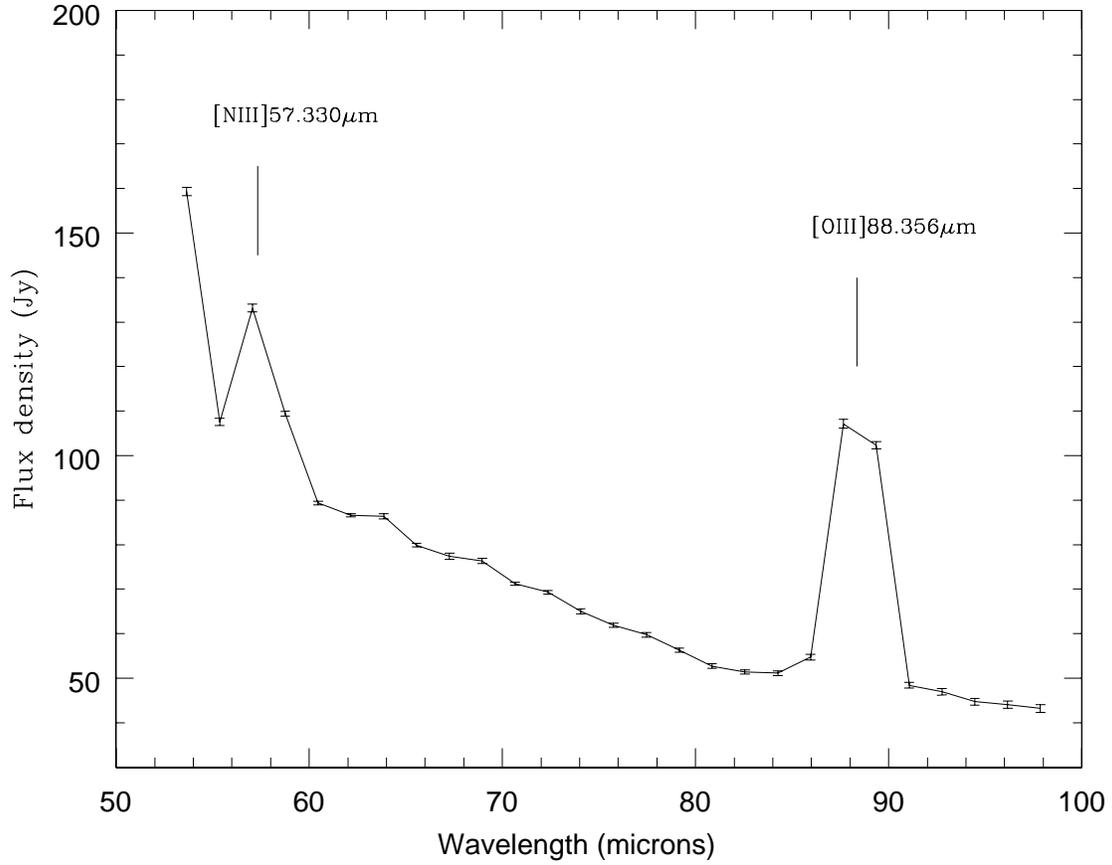}
\vspace{-0.7in}
\caption{
       The average SED-mode spectrum of the planetary nebula NGC$\,$6543 derived
       from observations made in MIPS Campaigns 20 and 36.  The spectrum is extracted 
       with a 5-column ($\sim$50\arcsec\/) aperture.
     }
\end{figure}

\begin{figure}
\plotone{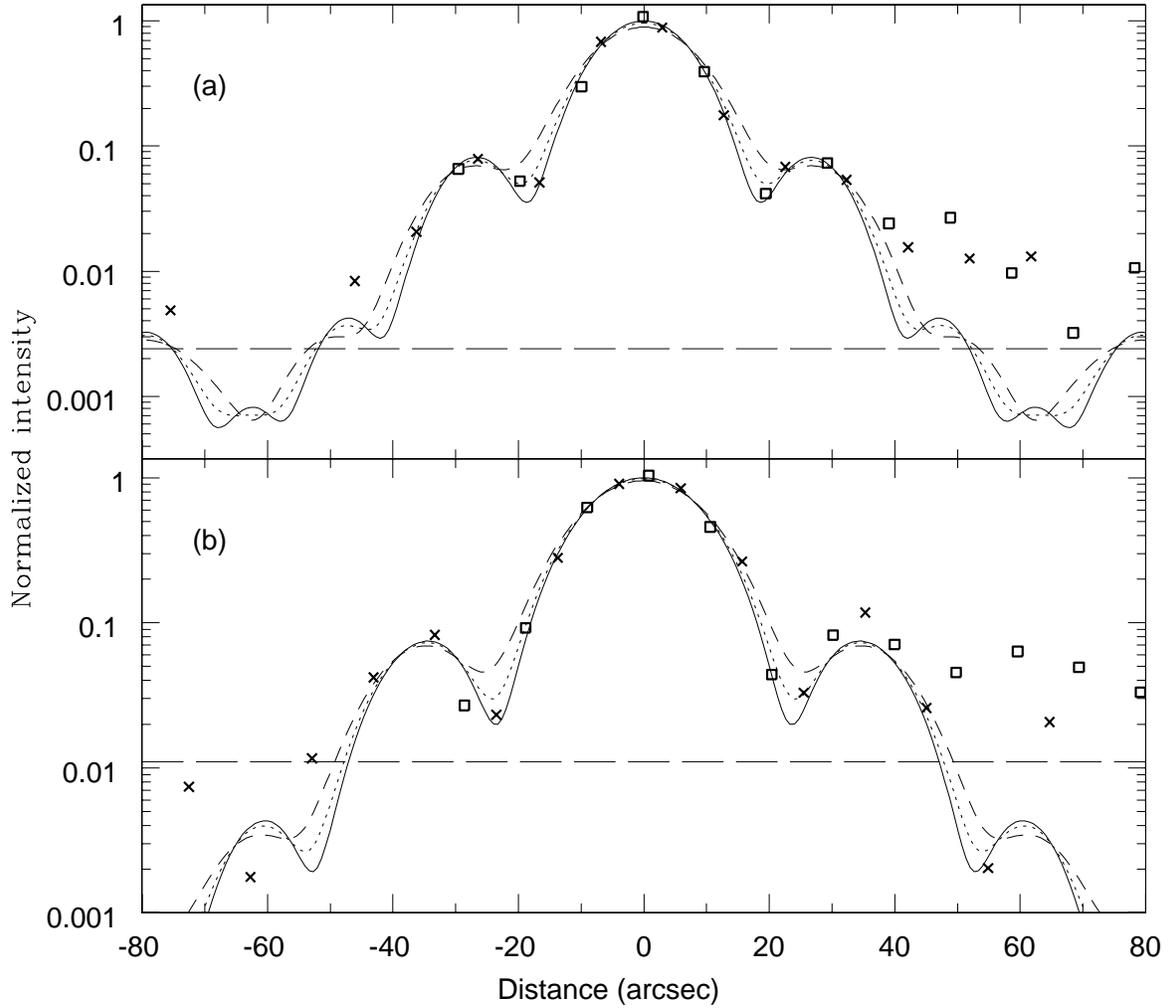}
\vspace{-0.4in}
\caption{Comparisons of the observed mean spatial profiles (points) with the model
	   predictions (curves) at 70~\micron\ (a) and 90~\micron\ (b).   The measured profiles 
	   are shown separately for dither positions 1 (crosses) and 2 (squares) and
           are derived from the observations of the bright calibration stars.
	   The model curves are those of the \Spitzer\ Tiny Tim model smoothed 
           by a boxcar of width of D $= 1.0$ (solid), $1.3$ (dotted) or 
	   $1.7$ (dashed) times the detector pixel size of 9.8\arcsec.
	}
\end{figure}

\begin{figure}
\plotone{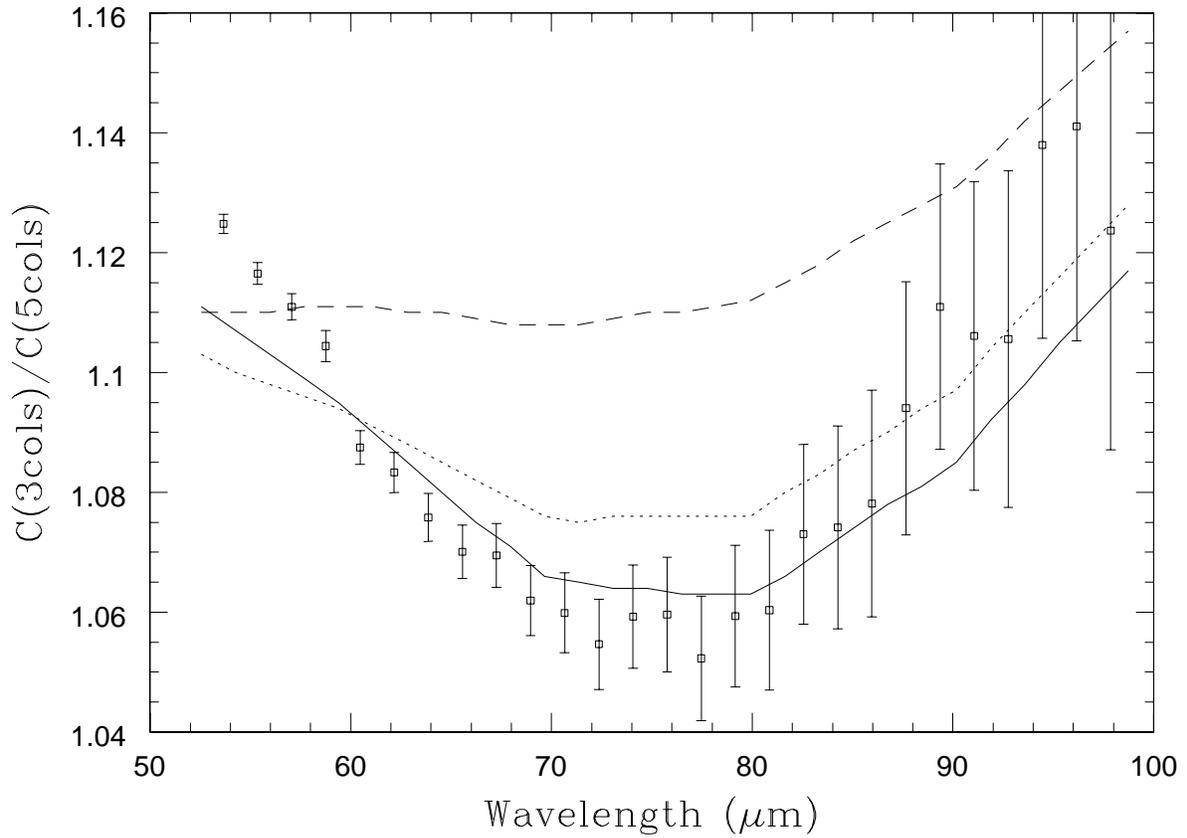}
\vspace{-0.6in}
\caption{Comparison between the observed (squares) and
Tiny Tim model predictions (curves) 
	 of the 3-column to 5-column aperture correction ratio, 
	 C(3cols)/C(5cols).
          The 3 model results shown here are denoted in the same manner as in 
          Fig.~5.
	}
\end{figure}

\begin{figure}
\plotone{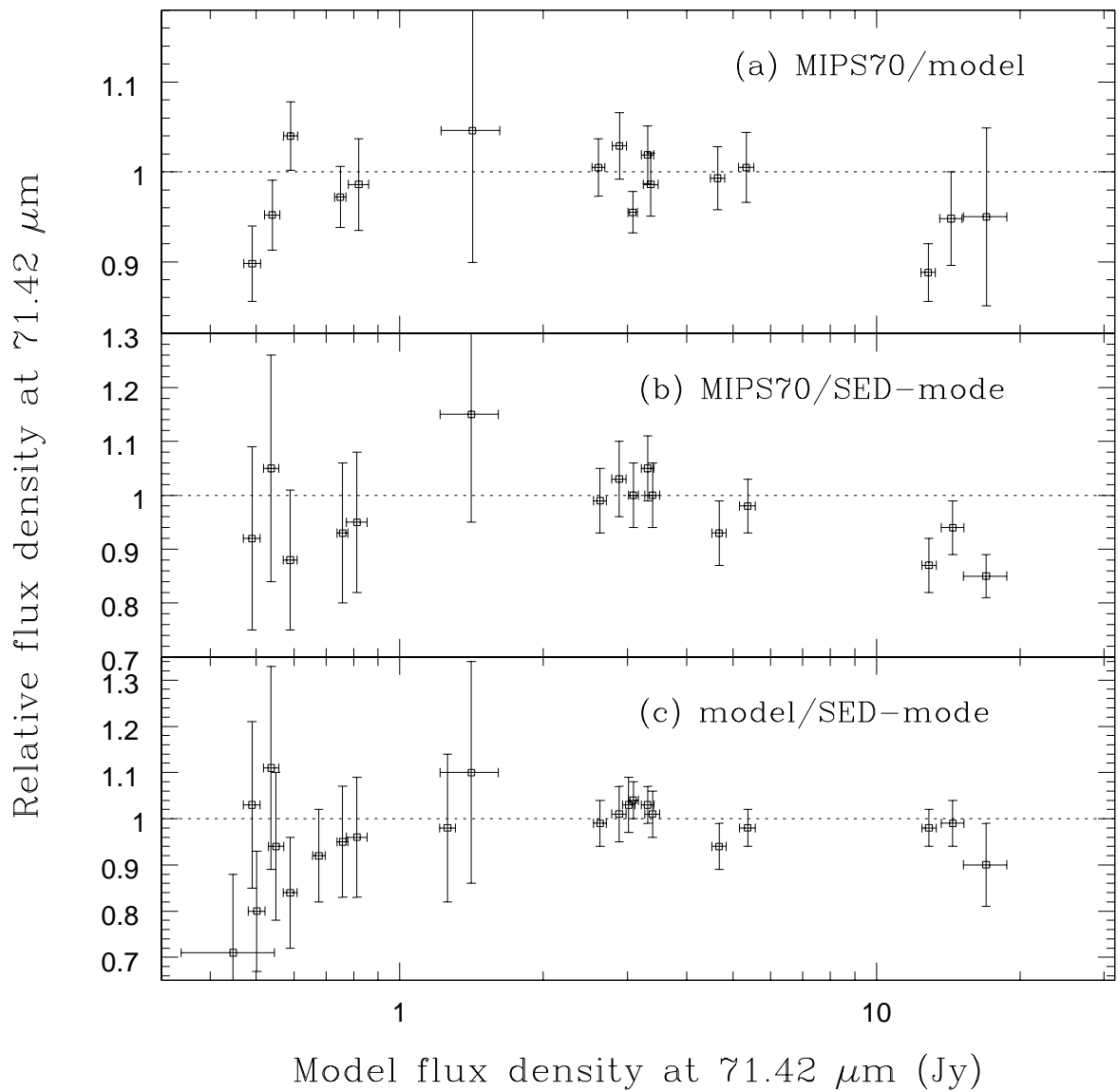}
\caption{
     Various  71.42~\micron\ flux density ratios as a function of the model predicted 
     flux density at 71.42~\micron. Panels (a) and (b) display the MIPS 70~\micron\ photometric 
     measurement relative to the model prediction and the SED-mode measurement, respectively. 
     Panel (c) shows the ratio of the model prediction to the SED-mode measurement.  
     The (relative) SED-mode flux densities were derived using a conversion factor 
     that yields a sample median for (c) close to 1 for those stars with $f_{\nu}(71.42)
     > 1$~Jy.
     }
\end{figure}

\begin{figure}
\plotone{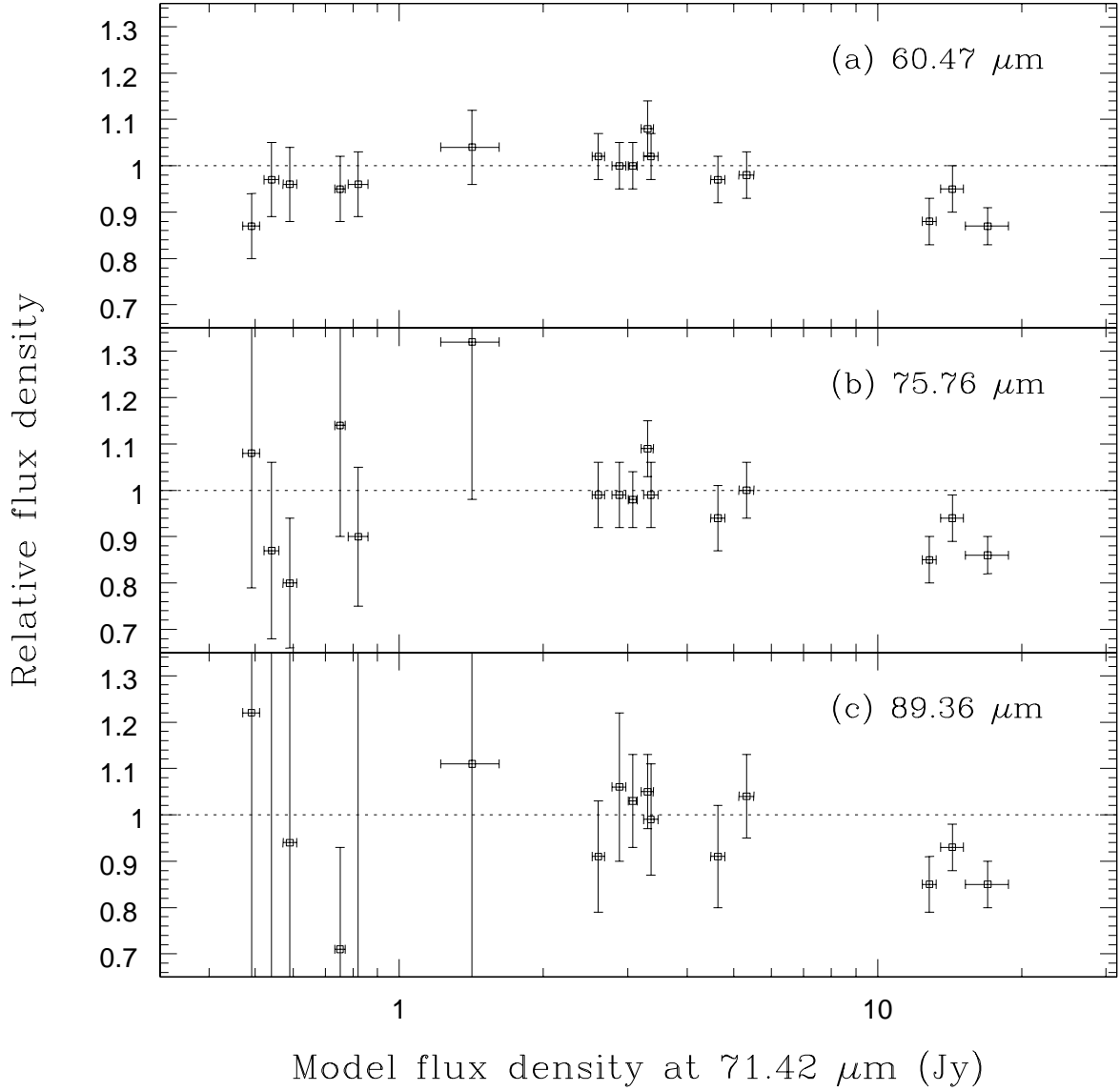}
\caption{
     The ratio of the predicted MIPS
     photometric flux densities to the SED-mode flux densities at wavelengths
of (a) 60.46~\micron, 
     (b) 75.76~\micron\ or (c) 89.36~\micron, as a function of the 71.42~\micron\ 
     model flux density.  
     }
\end{figure}

\begin{figure}
\plotone{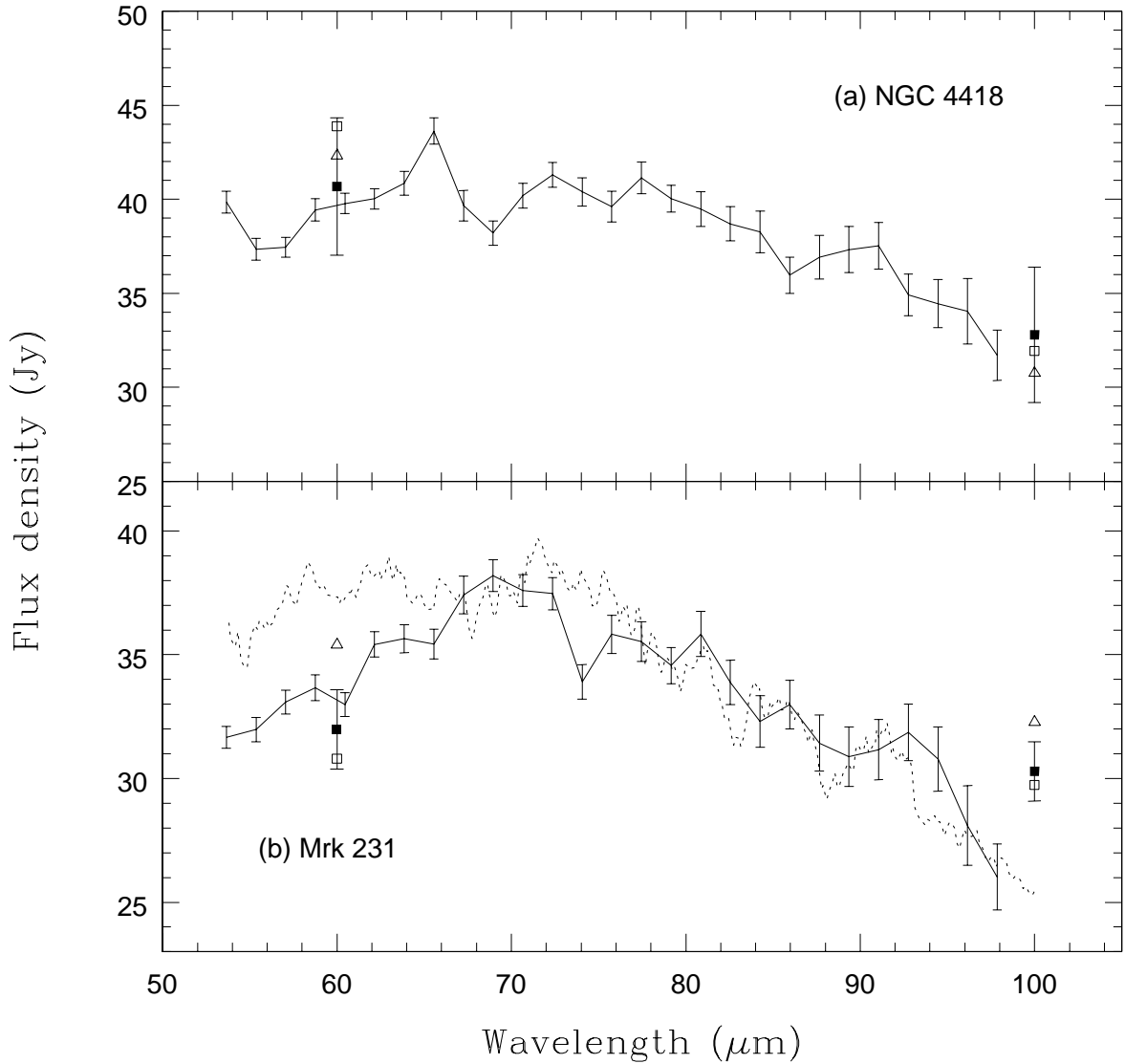}
\caption{Comparisons of the SED-mode measurements (solid lines) with \IRAS\ 60~\micron\ 
   and 100~\micron\ observations (squares and triangles) for the galaxies NGC$\,$4418 
   (a) and Mrk$\,$231 (b).  For the latter, an \ISO\ LWS spectrum (dotted line), 
   obtained from the \ISO\ archive (observation identifier 18001306) and smoothed using 
   a 7-pixel box car, is also shown. 
   Multiple sources for the \IRAS\ data are shown: IRAS Faint Source Catalog (solid
   squares), Sanders \etal (2003; open squares) and Soifer \etal (1989; open triangles).
   The \IRAS\ flux uncertainties are plotted only for the solid squares.
  }
\end{figure}

\begin{figure}
\plotone{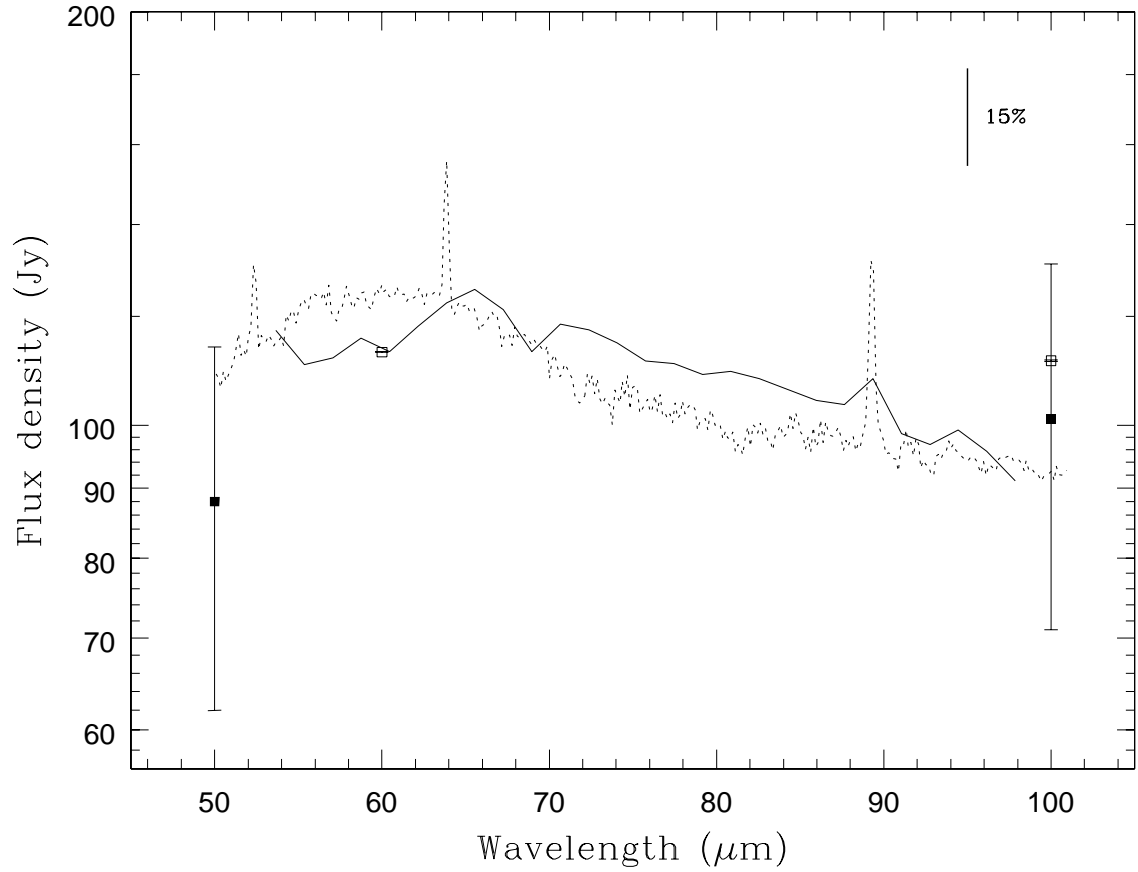}
\vspace{-0.4in}
\caption{Comparison of the spatially integrated SED-mode spectrum (solid line) with the \ISO\ LWS 
	 spectrum (dotted line) for the extended galaxy system Arp$\,$299.
	 The LWS 
         spectrum is derived from an \ISO\ archival observation (identifier 18001306).
         Also shown are measurements from \IRAS\ (Sanders \etal 2003;
         open squares) and Kuiper Airborne Observatory (Joy \etal 1989; solid squares).  
     }
\end{figure}

\end{document}